\documentclass[aps,twocolumn,superscriptaddress]{revtex4-2}

\usepackage{amsmath,amssymb,color,graphicx,mathpazo,mathtools,times,float,dsfont,bbm}
\usepackage[colorlinks=true,allcolors=blue]{hyperref}

\newcommand{\eref}[1]{Eq.~\eqref{eq:#1}}

\newcommand{\Fref}[1]{Figure~\ref{fig:#1}}
\newcommand{\fref}[1]{Fig.~\ref{fig:#1}}

\newcommand{\aref}[1]{App.~\ref{sec:#1}}
\newcommand{\idop}{\mathbbm{1}}

\begin{document}
\title{Routing entanglement through quantum networks}

\author{Karl Pelka}
\email{karl.a.pelka@um.edu.mt}
\affiliation{Department of Physics, University of Malta, Msida MSD 2080, Malta}
\author{Matteo Aquilina}
\email{matteo.aquilina.07@um.edu.mt}
\affiliation{Department of Physics, University of Malta, Msida MSD 2080, Malta}
\author{Andr\'e Xuereb}
\email{andre.xuereb@um.edu.mt}
\affiliation{Department of Physics, University of Malta, Msida MSD 2080, Malta}
\date{\today}

\begin{abstract}
Entanglement, one of the clearest manifestations of non-classical physics, holds significant promise for technological applications such as more secure communications and faster computations. In this paper we explore the use of non-reciprocal transport in a network of continuous-variable systems to route entanglement in one direction through the network. We develop the theory and discuss a potential realization of controllable flow of entanglement in quantum systems; our method employs only Gaussian interactions and engineered dissipation to break the symmetry. We also explore the conditions under which thermal fluctuations limit the distance over which the entanglement propagates and observe a counter-intuitive behavior between this distance, the strength of the entanglement source, and the strength of the hopping through the network.
\end{abstract}

\maketitle

\section{Introduction}
Applications of quantum technologies~\cite{Mohseni2017} rely on the exploitation of effects such as interference and entanglement~\cite{Horodecki2009} that are hallmarks of quantum mechanics. The potential that lies in entangled systems---correlations that surpass, and cannot be explained, by classical physics---is so promising that studies of networks~\cite{Acin2007} envisioning a quantum internet~\cite{Kimble2008} have created entire research fields~\cite{Xu2020,Pelucchi2022}. A useful resource for distributing entanglement in distant nodes is found in propagating photons~\cite{Cirac1997} enabling quantum key distribution through space using entanglement based schemes~\cite{Yin2020}. Here we explore the propagation of entanglement in a cascaded quantum system, showing that it is in principle possible to set up a scheme that allows entanglement to percolate one way, but not the other, through a network of quantum devices. Our exploration is phrased in the language of optomechanics and phonons, but can be applied to any set of interacting continuous-variable quantum systems.

The significance of non-reciprocity in quantum information applications is established in a variety of architectures, ranging from photonic setups~\cite{Kim2015,Dong2015} and superconducting circuits~\cite{Sliwa2015,Lecocq2017} to optomechanical systems~\cite{Shen2016,Bernier2017,Barzanjeh2017,Peterson2017,Mercier2019,Xu2019}. 
In a largely separate line of research, the dynamics of micro-mechanical systems subject to radiation pressure forces has been explored thoroughly within the field of optomechanics~\cite{Aspelmeyer2014,Barzanjeh2022}. This optomechanical interaction has also been used to explore the many-body dynamics of systems of coupled optomechanical networks, giving rise to phenomena including the possibility of obtaining stronger coupling at the single-photon level~\cite{Xuereb2012,Piergentili2018}, topological physics~\cite{Schmidt2015,Peano2015,Mathew2018}, dynamical gauge fields~\cite{Walter2016}, and synchronization~\cite{Heinrich2011,Lauter2017,Lipson2012,Holmes2012,Amitai2017,Loerch2017,Lauter2015,Lipson2015,Colombano2019}. Seminal work has shown the cooling of motion of the mechanical system down to its ground state~\cite{Chan2011,Teufel2011} and generation of quantum entanglement between mechanical oscillators~\cite{Ockeloen2018,Riedinger2018}. A recent focal point of interest within optomechanics lies in the research on nonreciprocal manipulation of fluctuations in quantum systems, including routing thermal noise with optomechanical thermal rectifiers~\cite{Barzanjeh2018} and swapping entanglement in hot modes using cold auxiliary modes~\cite{Orr2023}.

In this paper, we employ the vision of a quantum-communication network which consists of a number of nodes in a given geometry sharing some quantum correlations described by a quantum state $\rho$. We show that along a connected quantum channel a combination of coherent hopping and engineered dissipation enables or disables the flow of entanglement to another node achieving an entanglement routing device. Our results show that the creation of entanglement with our scheme may increase the effective temperature in the channel and, somewhat counter-intuitively, limit the propagation of entanglement through the system. Nevertheless, we find that entanglement can indeed propagate throughout a channel of several nodes and that state-of-the-art optomechanical experiments provide a basis that could realize our scheme.

We proceed by firstly deriving the effective quantum optics model based on cascaded quantum systems. This model will give us the physical intuition for the effects that occur and their limitations. Subsequently we employ an already established mapping to an optomechanical model where the mechanical degrees will replace the bath degrees and show that the mechanism can be realized in experiments with achievable parameters. Ultimately, we finish with a short outline on the implications of our results.

\section{Theoretical model}
We consider the collective dynamics of a squeezed bosonic mode coupled to a chain of $M$ bosonic modes acting as nodes in the transport channel described by the Hamiltonian

\begin{equation}
\hat{H}/\hbar=\sum\limits_{k=0}^{M}[\omega_k\hat{a}_{k}^{\dagger}\hat{a}_k+(iJ_k\hat{a}^{\dagger}_{k}\hat{a}_{k+1}+\text{h.c.})]+\frac{ir}{2}(\hat{a}_0^{\dagger 2}-\hat{a}_0^{2}),
\label{eq:Hamiltonian}
\end{equation}

with $\hat{a}_k$ describing the bosonic annihilation operators of the modes, $\omega_k$ their respective resonance frequencies, $J_k$ the hopping rates between neighboring nodes, and $r$ the squeezing parameter of the squeezed bosonic mode $\hat{a}_0\equiv\hat{a}_{M+1}$. Aiming to impose cascaded dynamics onto the transport channel, we follow Ref. \cite{Gardiner2000} by establishing common heat baths connecting neighboring oscillators $\hat{a}_l$ and $\hat{a}_{l+1}$. The oscillators decay with the rate $\gamma$ into their common bath with average occupation $\bar{N}_{l,l+1}$, realized with the standard dissipator in Lindblad form $\mathcal{D}_{\hat{o}}[\rho]=\hat{o}\rho\hat{o}^{\dagger}-\frac{1}{2}\{\rho,\hat{o}^{\dagger}\hat{o}\}$ through the collective operators $\hat{c}_{l}=(\hat{a}_l+\hat{a}_{l+1})/\sqrt{2}$ obeying the bosonic commutation relation $[\hat{c}_{l},\hat{c}^{\dagger}_{l}]=1$. The dynamics of the systems density matrix $\rho$ are described by the master equation in Lindblad form $\dot{\rho}=-(i/\hbar)[\hat{H},\rho]+\sum_{l=1}^{M-1}\gamma\{(\bar{N}_{l,l+1}+1)\mathcal{D}_{\hat{c}_l}[\rho]+\bar{N}_{l,l+1}\mathcal{D}_{\hat{c}^{\dagger}_l}[\rho]\}$ and the nonreciprocal effect can be achieved through the matching condition $J_l=\gamma/2$ for $l \in (1,...,M-1)$. The physical interpretation of the nonreciprocal effect is based on the interference of the direct hopping term from an oscillator to the next and the indirect path of hopping to the same oscillator through the common bath. The non-reciprocity relies on the coherent addition of these two paths in the direction $k \rightarrow k+1$ whereas the two terms cancel each other in the opposite direction $k \rightarrow k-1$. To allow for more generality, we admit decay with the rate $\gamma_{\text{out}}$ into a distinct bath with average occupation $\bar{N}_k$ for each of the oscillators $\hat{a}_k$ and obtain the master equation
\begin{multline}
\dot{\rho}=-\frac{i}{\hbar}[\hat{H},\rho] + \sum_{k=0}^{M}\gamma_{\text{out}}\{(\bar{N}_{k}+1)\mathcal{D}_{\hat{a}_k}[\rho]+\bar{N}_{k}\mathcal{D}_{\hat{a}^{\dagger}_k}[\rho]\}\\
+\sum_{j=1}^{M-1}\gamma\{(\bar{N}_{j,j+1}+1)\mathcal{D}_{\hat{c}_j}[\rho]+\bar{N}_{j,j+1}\mathcal{D}_{\hat{c}^{\dagger}_j}[\rho]\}.
\label{MasterEQN}
\end{multline}

\begin{figure}
\begin{center}
\includegraphics[width=0.9\columnwidth]{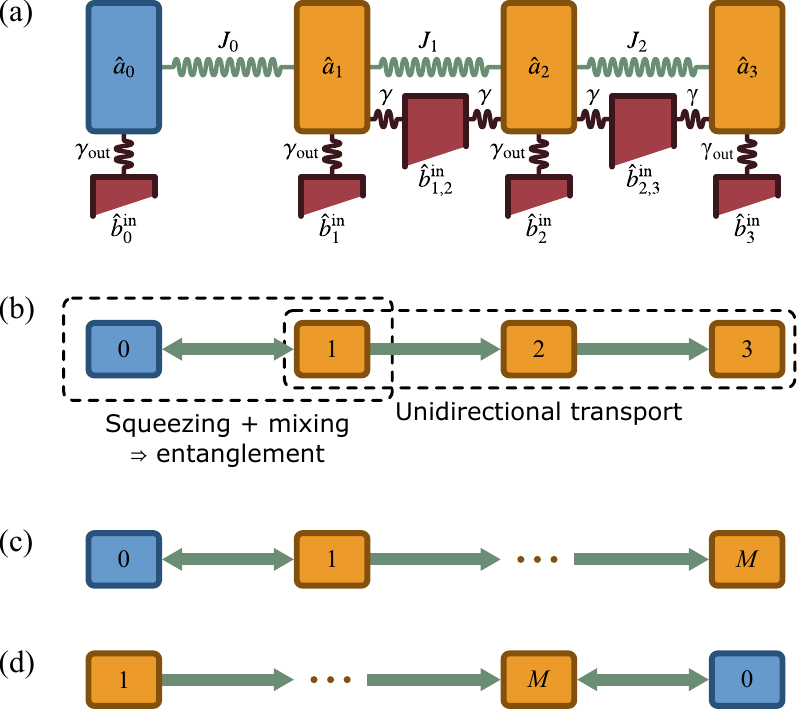}
\end{center}
\caption{(Color online) Schematic illustrations of a squeezed oscillator connected to chains consisting of unidirectionally coupled nodes. (a)~The squeezed node $\hat{a}_0$ is connected directly to one end of the chain constituting the quantum channel and the constituents of the channel are connected to their neighbors directly with rates $J_k$ as well as indirectly through a common bath in which the oscillators can decay excitations at rates $\gamma$. All nodes can decay excitations into distinct baths at rates $\gamma_{\text{out}}$. (b)~Abstract depiction of this chain and the directionality of the connections in this system. (c)~Abstraction of the node $\hat{a}_0$ connected to a quantum channel of $M$ nodes in the forward direction. (d)~Depiction of the node $\hat{a}_0$ connected to a quantum channel of $M$ nodes in the backward direction.}
\label{fig:Model}
\end{figure}

\Fref{Model} depicts a schematic illustration of the system of interest. The first panel, \fref{Model}(a), shows a small-scale model consisting of a squeezed mode $\hat{a}_0$ and three modes $\hat{a}_1$ to $\hat{a}_3$ that constitute a prototypical unidirectional chain. These modes are coupled directly with their nearest neighbors through hopping terms with rate $J_k$ in the Hamiltonian in \eref{Hamiltonian}. Each of the modes is coupled to a bath into which its excitations can decay at a rate $\gamma_{\text{out}}$. In addition, the nodes that constitute the unidirectional chain are pairwise connected through a common bath into which the connected modes can decay at the rate $\gamma$. \Fref{Model}(b) presents an abstract representation of this model by reducing nodes to vertices and showing directionality between vertices by arrows. \Fref{Model}(c) generalizes this model to any number $M$ of mode in the chain, whereas \fref{Model}(d) effectively reverses the direction of propagation in the chain by connecting the squeezed mode to the opposite end of the chain.

We proceed by converting the master equation from Eq. \eqref{MasterEQN} into equivalent quantum Langevin equations. Therefore, we derive the operator equations and use the fluctuation-dissipation theorem to add noise operators for the distinct baths $\hat{b}^{\text{in}}_{k}$ which obey $\langle (\hat{b}^{\text{in}}_{k})^{\dagger}(t)\hat{b}_{m}^{\text{in}}(t')\rangle=\bar{N}_k\delta_{km}\delta(t-t')$, $\langle \hat{b}^{\text{in}}_{k}(t)(\hat{b}^{\text{in}}_m)^{\dagger} (t') \rangle=(\bar{N}_k+1)\delta_{km}\delta(t-t')$ and for the common baths $\hat{b}^{\text{in}}_{l,l+1}$ obeying $\langle (\hat{b}_{l,l+1}^{\text{in}})^{\dagger}(t)\hat{b}_{m,m+1}^{\text{in}} (t') \rangle=\bar{N}_{l,l+1}\delta_{lm}\delta(t-t')$, $\langle \hat{b}^{\text{in}}_{l,l+1}(t)(\hat{b}^{\text{in}}_{m,m+1})^{\dagger} (t') \rangle=(\bar{N}_{l,l+1}+1)\delta_{l,m}\delta(t-t')$ while all other combinations vanish under the white-noise assumption. The resulting equations for the corresponding quadrature operators $\hat{x}_{\hat{o}}=(\hat{o}+\hat{o}^{\dagger})/\sqrt{2}$ and $\hat{p}_{\hat{o}}=-i(\hat{o}-\hat{o}^{\dagger})/\sqrt{2}$ can be summarized as $\dot{\vec{r}}=\mathcal{A}\vec{r}+\mathcal{B}\vec{r}_{\text{in}}$ for the quadrature vector $\vec{r}=(\hat{x}_{\hat{a}_0},\hat{p}_{\hat{a}_0},\dots,\hat{x}_{\hat{a}_M},\hat{p}_{\hat{a}_M})^T$ and noise quadrature vector $\vec{r}_{\text{in}}=(\hat{x}_{\hat{b}^{\text{in}}_{0}},\hat{p}_{\hat{b}^{\text{in}}_{0}},\dots,\hat{x}_{\hat{b}^{\text{in}}_{M}},\hat{p}_{\hat{b}^{\text{in}}_{M}},\hat{x}_{\hat{b}^{\text{in}}_{1,2}},\hat{p}_{\hat{b}^{\text{in}}_{1,2}},\dots,\hat{x}_{\hat{b}^{\text{in}}_{M-1,M}},\hat{p}_{\hat{b}^{\text{in}}_{M-1,M}})^T$ in terms of a dynamical matrix $\mathcal{A} \in \mathbb{R}^{2(M+1) \times 2(M+1)}$ and an adjacency matrix $\mathcal{B} \in \mathbb{R}^{2(M+1) \times 4M}$ which are documented in \aref{AppA}. Since the equations are linear, the state of the system at any point in time is fully described by the first $\langle \vec{r}_j \rangle$ and second moments of the quadrature operators. The second moments can be arranged into a covariance matrix $\mathcal{V}_{ij}=\langle \vec{r}_i \vec{r}_j + \vec{r}_j \vec{r}_i \rangle -2 \langle \vec{r}_i \rangle \langle \vec{r}_j \rangle$. Its dynamics can be shown to obey the Lyapunov equation $\dot{\mathcal{V}}=\mathcal{V}\mathcal{A}^{T}+\mathcal{A}\mathcal{V}+\mathcal{N}$ with the noise matrix $\mathcal{N}$ having elements $\mathcal{N}_{ij}=\sum_q(\bar{N}_{\tilde{q}}+\tfrac12)(\mathcal{B}_{iq}\mathcal{B}^{T}_{qj}+\mathcal{B}_{jq}\mathcal{B}^{T}_{qi})$, where $\bar{N}_{\tilde{q}}$ is the thermal occupancy of the bath referred to in the $q$th entry in $\vec{r}_{\text{in}}$.

Since the Hamiltonian is quadratic and the dissipators are of Lindblad form with respect to linear combinations of the annihilation and creation operators, the time evolution of the system is closed in the set of Gaussian states~\cite{Linowski2022}. Furthermore, it is also known that if the dynamics of the system is stable, i.e., the real parts of all the eigenvalues of $\mathcal{A}$ are negative~\cite{Behr2019}, then the dynamics governed by the Lyapunov equation reaches a unique steady state that can be obtained by solving the equation $\dot{\mathcal{V}}=0$ which is outlined in \aref{AppB}.

\section{Non-reciprocity of entanglement propagation}
These results allow us to our center of attention on the asymmetry of physical quantities between modes $k$ and $m$. Thus, we denote $\mathcal{V}^{(k,m)}$ as the reduced covariance matrix that characterizes the subsystem consisting of nodes $k$ and $m$ by ignoring all other entries~\cite{Serafini2017}. Accordingly, it can be further reduced to $2\times2$ blocks $\mathcal{V}^{(k)}$ and $\mathcal{V}^{(m)}$ which describe the covariance matrix of the respective single modes and contains the $2\times2$ correlation blocks $\mathcal{V}^{(k,m)}$. For the covariance matrix of a two-mode Gaussian state, the symplectic eigenvalues after partial transposition $\tilde{\nu}_i$ of the covariance matrix such as $\mathcal{V}^{(k,m)}$ allow to judge the separability of the subsystem~\cite{Adesso2004,Adesso2005,Serafini2004}. These eigenvalues can be computed with $\Delta(\mathcal{V}^{(k,m)})=\det(\mathcal{V}^{(k)})+\det(\mathcal{V}^{(m)})-2\det(\mathcal{V}^{(km)})$ as $\tilde{\nu}^{(k,m)}_\pm=\sqrt{\Delta(\mathcal{V}^{(k,m)})\pm\sqrt{\Delta(\mathcal{V}^{(k,m)})-4\det(\mathcal{V}^{(k,m)})}}/\sqrt{2}$. The PPT criterion for separability of the state characterized by $\mathcal{V}^{(k,m)}$ can be stated as $\tilde{\nu}^{(k,m)}_-\ge\tfrac12$, with the amount of entanglement being quantified through the logarithmic negativity~\cite{Adesso2004,Adesso2005,Serafini2004}
\begin{equation}
E^{(k,m)}_{\mathcal{N}}=\max\{0,-\ln(2\lvert\tilde{\nu}^{(k,m)}_-\rvert)\}.
\end{equation}

Our analysis proceeds with the properties of the steady state covariance matrix $\tilde{\mathcal{V}}$, given through the solution of $\tilde{\mathcal{V}}\mathcal{A}^{T}+\mathcal{A}\tilde{\mathcal{V}}+\mathcal{N}=0$ as the fixed point of the Lyapunov equation~\cite{Pietikainen2020}. The system under consideration consists of identical oscillators $\omega_k=\omega$ and a channel of length $M=10$ with $\gamma/\omega=0.8$, $\gamma_{\text{out}}/\omega=0.002$, and initial mean occupations $\bar{N}_k=\bar{N}_{l,l+1}=0$. A comment is in order here regarding the counter-intuitive nature of the value of $\gamma/\omega$. Perfect directionality in the chain can only be achieved if the hopping rates $J_l$ are equal to $\gamma/2$, as detailed above. Should the dissipation rate $\gamma$ be too small, the hopping rate between adjacent nodes would be similarly small and the entanglement would not propagate effectively through the chain. Conversely, if $\gamma$ is too large, dissipation is too strong and entanglement once again fails to propagate. The values of $\gamma$ and $J_l$ employed here strike a balance between these two scenarios and allow for the efficient propagation of entanglement. Further elaboration of the parameters is given in \aref{AppC}.

The behavior of the entanglement in the system is determined by the amount of squeezing of the auxiliary oscillator $\hat{a}_0$, tunable through the parameter $r$, and its connection to the chain with regards to the coupling structure's forward direction, as characterized by $J_{0}$ and $J_M$. The connection of the squeezed oscillator in the forward direction of the chain is realized by $iJ_0=j$ and $J_M=0$ while the connection against the forward direction of the chain requires $J_0=0$ and $-iJ_M=j$. We characterize the entanglement properties of the nonreciprocal coupling scheme through properties between the squeezed oscillator and the oscillator after the first nonreciprocal edge. \Fref{Nonreciprocity} illustrates the logarithmic negativity $E^{(0,\pm2)}_{\mathcal{N}}$ between oscillator $\hat{a}_0$ and the oscillator two hops away, with regards to the strength of squeezing and its coupling to the channel; in other words, $E^{(0,+2)}_{\mathcal{N}}:=E^{(0,2)}_{\mathcal{N}}$ quantifies the entanglement between the squeezed node its next-to-nearest neighbor in the forward direction, and $E^{(0,-2)}_{\mathcal{N}}:=E^{(0,9)}_{\mathcal{N}}$ in the reverse direction. We find that coupling the squeezed oscillator to the channel in the forward direction, depicted by the solid surface, enables nonzero values of the logarithmic negativity and we therefore find that the corresponding nodes are entangled. Moreover, we find that both squeezing and coupling are required to achieve entanglement as squeezing creates identical excitation pairs in $\hat{a}_0$ which can only travel to the channel if excitations can hop to the channel's first node and increasing both parameters results in a larger logarithmic negativity. If the squeezed oscillator is coupled to the channel against the forward direction as illustrated by the meshed surface, the resulting state between the channel and the squeezed oscillator is separable, evidenced by vanishing logarithmic negativity irrespective of the squeezing and coupling parameters. For large squeezing and hopping parameters the system is found to be unstable which can be seen in the edges of the surfaces resulting from the logarithmic negativity being undefined. Nevertheless, we find that entanglement of the constituents of the quantum channel only propagates in the forward direction of the unidirectional coupling.

\begin{figure}
 \centering
 \includegraphics[width=0.9\columnwidth]{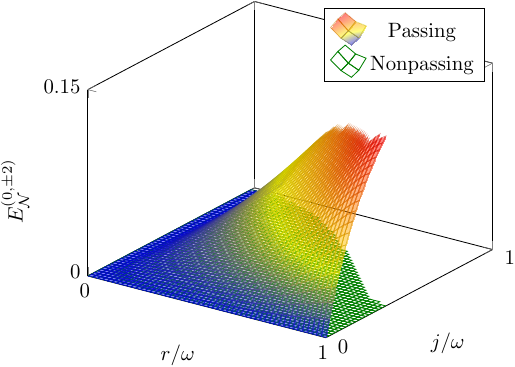}
 \caption{(Color online) Logarithmic negativity between the squeezed node and the node after the first nonreciprocal coupling in the forward direction (solid surface) and in the backward direction (meshed surface). The solid surface shows that connecting the squeezed oscillator to the channel in the forward direction results in nonzero logarithmic negativity indicating entanglement between the squeezed node and the nodes of the channel. The meshed surface shows that connecting the squeezed node against the forward direction results in zero logarithmic negativity irrespective of the strength of squeezing and coupling. Consequently, the squeezed oscillator and the channel are in a separable state demonstrating the effective isolation of entanglement propagation through the nonreciprocal coupling.}
\label{fig:Nonreciprocity}
\end{figure}

\begin{figure*}
\centering
\includegraphics[width=0.9\textwidth]{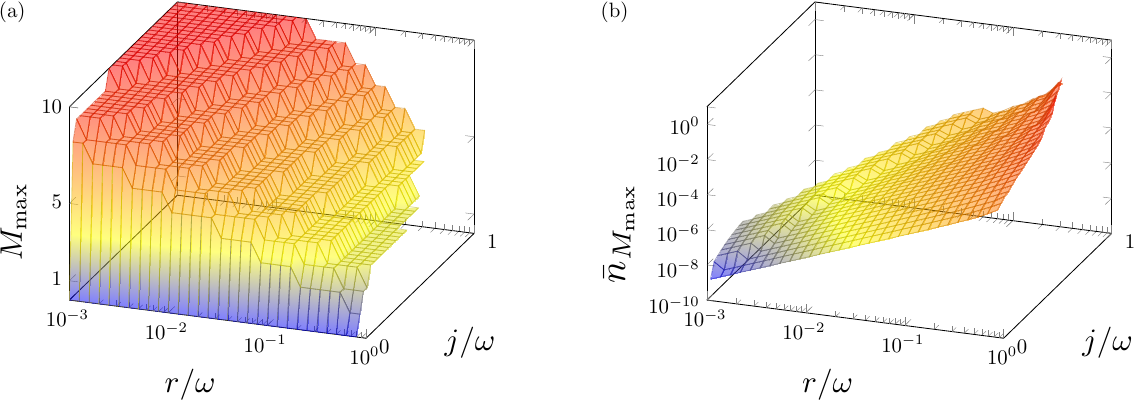}
\caption{(Color online) Propagation depth of entanglement and mean occupation number of the final entangled node in the quantum channel for the steady state in the forward direction. (a)~Maximal node in the quantum channel that is entangled with the squeezed oscillator depending on hopping and squeezing parameter. The maximal node increases with increasing the coupling of the squeezed oscillator to the quantum channel and decreases with the squeezing. For low squeezing and large hopping entanglement passes throughout the entire system. Increasing the squeezing leads to instability of the system. (b)~The mean occupation number of the maximal entangled node is related to the temperature increase generated through the entangling scheme. The mean occupation increases with larger coupling of the squeezed oscillator to the quantum channel as well as with the squeezing until the system becomes unstable. This result suggests that entangling the nodes in the system with the suggested scheme increases the temperature in the system until the system becomes unstable.}
\label{fig:PDepth}
\end{figure*}

\section{Influence of thermal fluctuations}
Aiming to understand the amount of nodes that can be entangled through the channel and its limits, we find the maximal node $M_{\text{max}}$ inseparable from the squeezed node by the largest $M$ with logarithmic negativity $E^{(0,M)}_{\mathcal{N}}>0$. The results of this analysis are depicted in \fref{PDepth}(a). Our results show that an increasing coherent coupling of the squeezed oscillator to the chain increases the amount of entangled nodes in the chain. We explain this increase by the improvement of correlated excitations to propagate to the chain with an enlarged hopping rate. However, we find that a stronger squeezing and therefore a larger rate of generating excitation pairs leads to fewer nodes that end up in an entangled state with the squeezed node and ultimately to an unstable system. Additionally, the single mode covariance matrices $\mathcal{V}^{(k)}(t)$ are related to the mean occupation of each node as $\bar{N}_{k}(t)=[\text{Tr}(\mathcal{V}^{(k)}(t))-2]/2$ which enables us to characterize the thermal noise that is generated in the respective node. In the following, we consider the mean occupation $\bar{n}_{k}=[\text{Tr}(\tilde{\mathcal{V}}^{(k)})-2]/2$ using the steady state~\cite{Pietikainen2020}. We attribute this occupation to the thermal noise which is generated through the creation and propagation of entanglement throughout the quantum channel in our scheme. The resulting mean occupation in the maximal node inseparable from the squeezed node are depicted in \fref{PDepth}(b). We find that the mean occupation number and therefore the temperature in the node rises with an increasing hopping rate and with an increasing squeezing parameter. Ultimately, the figure shows that the system becomes unstable when the occupation number becomes too large. We thus conclude that the same mechanism that generates entanglement also increases the effective temperature in the system, which is the limiting factor in the propagation of entanglement through the system.

\section{Optomechanical implementation}
This section aims to employ prior results~\cite{Barzanjeh2018,Barzanjeh2015} showing that mechanical degrees of freedom can act as controllable reservoirs required for our entanglement scheme. Consequently, the optomechanical system can be used as an entanglement routing channel for the optical fields. We employ the mapping worked out in~\cite{Barzanjeh2018} which shows that the nonreciprocal coupling can be achieved with dielectric nanostring mechanical resonators~\cite{Barzanjeh2017} or on-chip-microwave electromechanical systems based on a lumped-element superconducting circuit with a drumhead capacitor~\cite{Peterson2017,Bernier2017}. The resulting parameters amount to cavities resonant at $2\pi\times5$~GHz with damping rates $\kappa=2\pi\times2$~MHz, mechanical resonance frequency $\omega_{\text{m}}=2\pi\times6$~MHz, and damping rate $\gamma_{\text{m}}=2\pi\times100$~Hz. Although the value of $\gamma$, to which the values of $J_l$ are tied, as discussed previously, is significantly higher than is the norm for optomechanical experiments, we believe that such values are not outside the realm of experimental realization for systems that are designed and optimized appropriately.

\section{Conclusion and outlook}
Our investigation outlined a general framework for nonreciprocal propagation of entanglement in composite bosonic quantum systems. Our study also shows that the creation of entanglement increases the temperature in the system which in turn is the limiting factor for the stability of the system. Our framework can be mapped to an optomechanical realization. This mapping shows that the scheme can be realized with parameters of typical microwave optomechanical experiments. Therefore, our scheme can be tested in present day experiments with state-of-the-art setups in the optical and the microwave domain. In the context of quantum measurements and emerging quantum technologies, our techniques and ideas will find use in the manipulation of flow of entanglement inside quantum devices for phonon-based signal processing and computation.

\begin{acknowledgments}
The authors thank Shabir Barzanjeh for insightful discussions and feedback. KP and AX acknowledge financial support from the Ministry for Education, Sports, Youths, Research and Innovation of the Government of Malta through its participation in the QuantERA ERA-NET Cofund in Quantum Technologies (project MQSens) implemented within the European Union’s Horizon 2020 Programme.
\end{acknowledgments}

\appendix

\section{The dynamical and noise matrix in the Lyapunov equation}
\label{sec:AppA}
The unidirectionally coupled nodes described in \cite{Gardiner2000} were used as a basis for the model in this study and can be summarized through the master equation in Eq. (\ref{MasterEQN}). The equivalent quantum Langevin equations for the model in this study are given by
\begin{align}
\dot{\hat{a}}_k=&-\bigg(i\omega_k-\frac{\gamma_{\text{out}}+(2-2\delta_{k0}-\delta_{k1}-\delta_{kM})\gamma}{2}\bigg) \hat{a}_k \nonumber \\
&+\delta_{k0}(-r\hat{a}_0^{\dagger}+J_0\hat{a}_{1})-\delta_{k1}J_0^*\hat{a}_{0} \nonumber \\
&-(1-\delta_{k0}-\delta_{k1})\gamma\hat{a}_{k-1}+(1-\delta_{k0}-\delta_{kM})\sqrt{\gamma}\hat{b}^{\text{in}}_{k,k+1} \nonumber \\
&+(1-\delta_{k0}-\delta_{k1})\sqrt{\gamma}\hat{b}^{\text{in}}_{k-1,k} +\sqrt{\gamma_{\text{out}}}\hat{b}^{\text{in}}_{k}
\end{align}
in the case of connecting the source node to the first node in the passing direction, where $iJ_0=j_{\text{fwd}}$, $J_M=0$, and $J_1=J_2=....=J_{M-1}=\gamma/2$ according to the condition required for unidirectionality. Similarly, in the case of connecting the source node to the final node to obtain a connection against the passing direction with $J_0=0$, $-iJ_M=j_{\text{bwd}}$ and $J_1=J_2=....=J_{M-1}=\gamma/2$ the resulting dynamics turn out to be
\begin{align}
\dot{\hat{a}}_k=&-\bigg(i\omega_k-\frac{\gamma_{\text{out}}+(2-2\delta_{k0}-\delta_{k1}-\delta_{kM})\gamma}{2}\bigg) \hat{a}_k \nonumber \\
&+\delta_{k0}(-r\hat{a}_0^{\dagger}-J_M\hat{a}_{M})+\delta_{kM}J_M^*\hat{a}_{0} \nonumber \\
&-(1-\delta_{k0}-\delta_{k1})\gamma\hat{a}_{k-1}+(1-\delta_{k0}-\delta_{kM})\sqrt{\gamma}\hat{b}^{\text{in}}_{k,k+1} \nonumber \\
&+(1-\delta_{k0}-\delta_{k1})\sqrt{\gamma}\hat{b}^{\text{in}}_{k-1,k}+\sqrt{\gamma_{\text{out}}}\hat{b}^{\text{in}}_{k}
\end{align}

The quantum Langevin equations define the dynamics of the system and can be rewritten in terms of the two quadratures $\hat{x}_{k}=(\hat{a}_{k}^\dagger+\hat{a}_{k})/\sqrt{2}$ and $\hat{p}_k=i(\hat{a}_k^\dagger-\hat{a}_k)\sqrt{2}$ of each mode as
\begin{align}
\frac{\text{d}}{\text{d}t}\hat{r}(t)=\mathcal{A}\vec{r}_l(t) + \mathcal{B}\vec{r}_{\text{in}}(t)
\end{align}
with the quadratures arranged as the vector $\vec{r}=(\hat{x}_{\hat{a}_0},\hat{p}_{\hat{a}_0},\dots,\hat{x}_{\hat{a}_M},\hat{p}_{\hat{a}_M})^{T}$ and the noise  vector $\vec{r}_{\text{in}}=(\hat{x}_{\hat{b}^{\text{in}}_{0}},\hat{p}_{\hat{b}^{\text{in}}_{0}},\dots,\hat{x}_{\hat{b}^{\text{in}}_{M}},\hat{p}_{\hat{b}^{\text{in}}_{M}},\hat{x}_{\hat{b}^{\text{in}}_{1,2}},\hat{p}_{\hat{b}^{\text{in}}_{1,2}},\dots,\hat{x}_{\hat{b}^{\text{in}}_{M-1,M}},\hat{p}_{\hat{b}^{\text{in}}_{M-1,M}})^{T}$
where $\mathcal{A}_{kl}$ denotes a $2M \times 2M$ matrix called the dynamical matrix. 

From the equations (S1) and (S2) we extract dynamical matrix $\mathcal{A}$ as it only encompasses terms containing the nodes $\hat{a}$. The dynamical matrix can be written with the $2 \times 2$ identity matrix $\idop_2$, the $2 \times 2$ zero matrix $0_2$, and the Pauli matrices
\begin{align}
\sigma_x=\begin{bmatrix} 0& 1 \\1& 0 \end{bmatrix}, \
\sigma_y=\begin{bmatrix} 0& -i \\i& 0 \end{bmatrix}, \
\sigma_z=\begin{bmatrix} 1& 0 \\0& -1 \end{bmatrix}
\end{align} as
\begin{widetext}
\begin{align}
\mathcal{A}=\begin{bmatrix}
-r\sigma_z-\frac{\gamma_{0}}{2}\idop_2+i\omega_0\sigma_y & i j_{\text{fwd}}\sigma_y&0_2& \hdots &i j_{\text{bwd}}\sigma_y\\ 
i j_{\text{fwd}}\sigma_y & -\frac{\gamma_{1}}{2}\idop_2+i\omega_1\sigma_y & 0_2 & \hdots & 0_2 \\
0_2 &-\gamma \idop_2 & -\frac{\gamma_{2}}{2}\idop_2+i\omega_2\sigma_y & \hdots & 0_2 \\
0_2 & 0_2 &-\gamma \idop_2 & \hdots & 0_2 \\
\vdots & \vdots & \vdots & \ddots & \vdots \\
i j_{\text{bwd}}\sigma_y & 0_2 & 0_2 & \hdots & -\frac{\gamma_{M}}{2}\idop_2+i\omega_M\sigma_y
\end{bmatrix}
\end{align}
\end{widetext}
with $\gamma_0=\gamma_{\text{out}}$, $\gamma_1=\gamma_M=\gamma_{\text{out}}+\gamma$ and all other $\gamma_k=\gamma_{\text{out}}+2\gamma$ according to the number of baths to which these are connected.
We can extract $\mathcal{B}$ from Eq. (1) and Eq. (2) as $2\times2$ identity matrices $\idop_{2}$ if the respective bath is connected to the oscillator scaled by the square root of the decay rate and zeros otherwise
\begin{widetext}
\begin{align}
\mathcal{B}=\begin{bmatrix}
\sqrt{\gamma_{\text{out}}}\idop_2&0_2&0_2&\hdots&0_2&0_2\hspace{0.1cm}&0_2&\hdots&0_2 \\
0_2&\sqrt{\gamma_{\text{out}}}\idop_2&0_2&\hdots&0_2&0_2\hspace{0.1cm}&\sqrt{\gamma}\idop_2&\hdots&0_2 \\
0_2&0_2&\sqrt{\gamma_{\text{out}}}\idop_2&\hdots&0_2&0_2\hspace{0.1cm}&\sqrt{\gamma}\idop_2&\hdots&0_2 \\
\vdots&\vdots&\vdots&\ddots&\vdots&\vdots\hspace{0.1cm}&\vdots&\ddots&\vdots \\
0_2&0_2&0_2&\hdots&\sqrt{\gamma_{\text{out}}}\idop_2&0_2\hspace{0.1cm}&0_2&\hdots&\sqrt{\gamma}\idop_2 \\
0_2&0_2&0_2&\hdots&0_2&\sqrt{\gamma_{\text{out}}}\idop_2\hspace{0.1cm}&0_2&\hdots&\sqrt{\gamma}\idop_2 \\
\end{bmatrix}
\end{align}
\end{widetext}
With the covariance matrix defined as
\begin{align}
\mathcal{V}_{ij}=&\langle\vec{r}_i\vec{r}_j+ \vec{r}_j\vec{r}_i\rangle-2\langle\vec{r}_i\rangle\langle\vec{r}_j\rangle
\end{align}
Through some manipulation the dynamics can be worked out in terms of the covariance matrix whereby the resulting equation is termed the Lyapunov equation: 
\begin{align}
\frac{\text{d}}{\text{d}t}\mathcal{V}=\mathcal{V}\mathcal{A}^{T}+\mathcal{A}\mathcal{V}+\mathcal{N}
\label{LyaEqn}
\end{align}
with the noise matrix 
\begin{align}
\mathcal{N}_{ij}=\sum_{q}(\bar{N}_{\lfloor q/2 \rfloor}+1/2)(\mathcal{B}_{iq}\mathcal{B}^{T}_{qj}+\mathcal{B}_{jq}\mathcal{B}^{T}_{qi})
\label{NoiseMatrix}
\end{align}
consisting of $2 \times 2$ blocks $\gamma_q(\bar{N}_q+1/2)\idop_2$ if oscillator $i$ and oscillator $j$ are coupled to the bath referred to in the $q$th index in $\vec{r}_{\text{in}}$.

\section{Stability and Physicality analysis of the system}
\label{sec:AppB}

\begin{figure}
\begin{center}
\includegraphics[width=0.5\textwidth,angle=0]{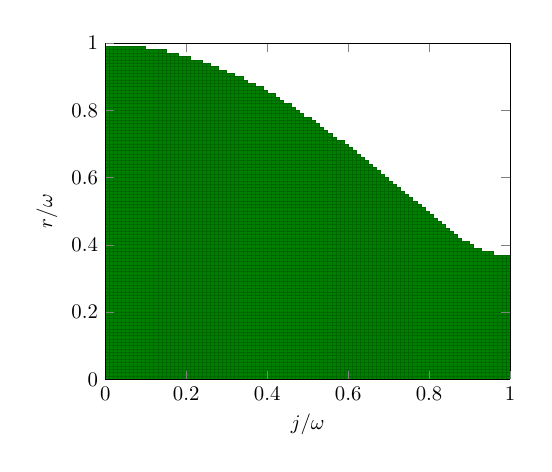}
\caption{Stability and physicality of the system depending on the coupling strength $j/\omega$ and squeezing parameter $r/\omega$ for the parameters $\gamma_{\text{out}}/\omega$=0.002 and $\gamma/\omega$=0.8 used in the analysis.}
\label{fig:F4}
\end{center}
\end{figure}

We have established that the covariance matrix evolves under the differential Lyapunov equation
\begin{align}
\frac{\text{d}}{\text{d}t}\mathcal{V}(t)=&\mathcal{A}\mathcal{V}(t)+\mathcal{V}(t)\mathcal{A}^{T}+\mathcal{N} \nonumber \\
\mathcal{V}(0)=&\mathcal{V}_0
\label{DLP}
\end{align}
where we assume that $\mathcal{A}$ is diagonalizable such that $P^{-1}\mathcal{A}P=D$ with the diagonal matrix $D$. Consequently, the transpose is $\mathcal{A}^T$ is diagonalizable as $Q^{-1}\mathcal{A}Q=D=P^{T}\mathcal{A}^T(P^{-1})^T=D^T=D$ with $Q=(P^T)^{-1}$. Then, it can be shown that the solution to the dynamical Lyapunov equation is unique and can be spectrally decomposed~\cite{Behr2019}
\begin{align}
\mathcal{V}(t)=&P\bigg[\bigg(\int\limits_{0}^{t}e^{(\alpha_j+\alpha_k)(t-s)}\text{d}s\bigg)_{jk}\circ (P^{-1}\mathcal{N} (Q^\dagger)^{-1})_{jk})\bigg]Q^\dagger
 \nonumber \\
 &+P[(e^{(\alpha_j+\alpha_k)t})_{jk}\circ (P^{-1}\mathcal{V}_0(Q^\dagger)^{-1})_{jk}]Q^\dagger,
 \label{SteadyState}
\end{align}
where $\circ$ denotes the Hadamard product or equivalently entry-wise multiplication of the matrices. Through the scalar integrals~\cite{Behr2019}
\begin{align}
\int\limits_{0}^{t}e^{(t-s)(\alpha_j+\alpha_k)}\text{d}s=\begin{cases} \frac{e^{(\alpha_j+\alpha_k)t}-1}{(\alpha_j+\alpha_k)}& \text{ if } (\alpha_j+\alpha_k)\ne 0 \\t & \text{ if } (\alpha_j+\alpha_k)= 0
\end{cases}
\end{align}
we see, that $\mathcal{V}$ acquires a unique steady state if all eigenvalues of $\mathcal{A}$ have negative real part. Then, the steady state for $t\rightarrow \infty$ is given by the first line of Eq. (\ref{SteadyState}) as the second line vanishes.\par
A test for physicality of the resulting covariance matrix consists in satisfying two requirements~\cite{Adesso2004}. Firstly, the steady state covariance matrix must be positive which means that the real parts of the eigenvalues of the covariance matrix are all equal to or greater than zero. Secondly, all symplectic eigenvalues of the covariance matrix must have real parts greater than or equal to $1/2$. The symplectic eigenvalues can be computed with the symplectic form
\begin{equation}
\Omega=\bigoplus_{i=0}^{M}\begin{pmatrix} 0 & 1 \\
-1 & 0
\end{pmatrix}
\end{equation}
as the eigenvalues of the matrix $|i\Omega\mathcal{V}|$.
The results for the physicality happen to coincide with those of stability throughout our analyses. Therefore, the resulting stability and physicality of the system depending on the coupling strength $j/\omega$ and squeezing parameter $r/\omega$ for the parameters used in the analysis are illustrated in \fref{F4}.

\section{Determination of the operational parameters}
\label{sec:AppC}
The matching between the cascaded quantum system with an optomechanical system as well as the aims of our analyses influence the various parameters of the model. We assumed that the mean number of excitations $\bar{N}_k=[exp(\hbar\omega/k_BT_k)-1]^{-1}$ is chosen to observe the dynamics of the optomechanical system without the influence of noise. Thus, we set $\bar{N}_k=0$ throughout. The resulting mean number of excitations in the system, depicted in \fref{PDepth}(b), can be converted into an effective temperature $T_{\text{eff}}$ generated through the process depending on the chosen frequency $\omega$ through inversion of the above equation as
\begin{align}
T_{\text{eff}}=\frac{\hbar \omega}{k_B \text{ln}\Big(1+\frac{1}{\bar{n}_{M_{\text{max}}}}\Big)}
\end{align}
Using the frequency $\omega_{\text{cav}}=2\pi \times 5$ GHz of the cavity suggested and a cut-off mean occupation of $\bar{n}_{M_{\text{max}}}=0.01$, we obtain an effective temperature of 52 mK that can be generated through the process which is much larger than the ambient temperature of $10$ mK at which the suggested microwave resonators can be operated~\cite{Barzanjeh2018}. Thus, the effects of the non-reciprocal entanglement routing should be observable in experiments and not be hidden by thermal noise.

\par
The values assigned for both squeezing and hopping terms, $r$ and $j$ respectively, are assumed to be in the small coupling region in which perturbative approaches apply, which limits these parameters to be bounded by the oscillator frequencies. Therefore, we can scale $r$ and $j$ accordingly in terms of $\omega$ and perform our analyses for $0 \le r/\omega,j/\omega \le 1$. Similarly to the temperature, the squeezing parameter $r$ employed here is always scaled to the frequency $\omega$ and thus the squeezing to be employed in an experimental realization depends on the frequency of the system under consideration. \par
The value of $\gamma$ was decided upon, from a bounded region identical to that of $r$ and $j$, after several observations throughout the study. The chosen value for $\gamma/\omega=0.8$ is observed to fall within a substantially narrow range where manifestation of entanglement is favoured. The coupling strength to the respective baths, $\gamma_{\text{out}}/\omega$, was observed to have a minimal impact on the system and thus its value was kept quite small, concretely $\gamma_{\text{out}}/\omega=0.002$, so that the study would be able to focus on the more critical parameters.

\end{document}